 \documentclass[final,5p,times,twocolumn,authoryear]{elsarticle}

\usepackage{amssymb}
\usepackage{lipsum}

\def\azh{\ref@jnl{AZh}}                 
\def\baas{\ref@jnl{BAAS}}               
\def\bac{\ref@jnl{Bull. astr. Inst. Czechosl.}}

\def\jcap{\ref@jnl{J. Cosmology Astropart. Phys.}}
\def\jrasc{\ref@jnl{JRASC}}             
\def\memras{\ref@jnl{MmRAS}}            
\def\nar{\ref@jnl{New A Rev.}}          
\def\pra{\ref@jnl{Phys.~Rev.~A}}        
\def\prb{\ref@jnl{Phys.~Rev.~B}}        
\def\prc{\ref@jnl{Phys.~Rev.~C}}        
\def\prd{\ref@jnl{Phys.~Rev.~D}}        
\def\pre{\ref@jnl{Phys.~Rev.~E}}        
\def\prl{\ref@jnl{Phys.~Rev.~Lett.}}    
\def\rmxaa{\ref@jnl{Rev. Mexicana Astron. Astrofis.}}%
\def\qjras{\ref@jnl{QJRAS}}             
\def\skytel{\ref@jnl{S\&T}}             
\def\solphys{\ref@jnl{Sol.~Phys.}}      
\def\sovast{\ref@jnl{Soviet~Ast.}}      
\def\ssr{\ref@jnl{Space~Sci.~Rev.}}     
\def\zap{\ref@jnl{ZAp}}                 
\def\iaucirc{\ref@jnl{IAU~Circ.}}       

\def\bain{\ref@jnl{Bull.~Astron.~Inst.~Netherlands}} 
\def\fcp{\ref@jnl{Fund.~Cosmic~Phys.}}  
\def\gca{\ref@jnl{Geochim.~Cosmochim.~Acta}}   
\def\grl{\ref@jnl{Geophys.~Res.~Lett.}} 
\def\jcp{\ref@jnl{J.~Chem.~Phys.}}      
\def\jgr{\ref@jnl{J.~Geophys.~Res.}}    
\def\jqsrt{\ref@jnl{J.~Quant.~Spec.~Radiat.~Transf.}}
\def\memsai{\ref@jnl{Mem.~Soc.~Astron.~Italiana}}
\def\nphysa{\ref@jnl{Nucl.~Phys.~A}}   
\def\physrep{\ref@jnl{Phys.~Rep.}}   
\def\physscr{\ref@jnl{Phys.~Scr}}   

\newcommand{\beq}{\begin{equation}}
\newcommand{\eeq}{\end{equation}}
\newcommand{\beqn}{\begin{eqnarray}}
\newcommand{\eeqn}{\end{eqnarray}}
\newcommand{\non}{\nonumber}
\newcommand{\no}{\noindent}

\journal{New Astronomy}

\begin{document}

\begin{frontmatter}



\title{Searching for Candidates of Orbital Decays among Transit Exoplanets}


\author[label1]{Li-Chin Yeh}
\author[label2,label3]{Ing-Guey Jiang}
\ead{jiang@phys.nthu.edu.tw}
\author[label2]{Napaporn A-thano}
\address[label1]{Institute of Computational and Modeling Science,
National Tsing-Hua University, Hsin-Chu, Taiwan}
\address[label2]{Department of Physics and Institute of Astronomy, National Tsing-Hua University, Hsin-Chu, Taiwan}
\address[label3]{Center for Informatics and Computation in Astronomy, National Tsing-Hua University, Hsin-Chu, Taiwan}

\begin{abstract}
Transit observations have become an important technique to probe exoplanets. Therefore, there are many projects carrying on organized observations 
of transit events, which make a huge amount of light-curve and transit timing data available. 
We consider this as an excellent opportunity to search for possible orbital decays 
of exoplanets from this big number of mid-transit times through data-model fitting with both fixed-orbit and orbit-decay models. 
In order to perform this task, 
we collect mid-transit-time data from several 
sources and construct the most complete database up to date.  Among 144 hot Jupiters in our study,
HAT-P-51b, HAT-P-53b, TrES-5b, WASP-12b
are classified as the orbit-decay cases.
Thus, in addition to reconfirming WASP-12b as an orbit-decay planet, our results indicate that 
HAT-P-51b, HAT-P-53b, TrES-5b are potential orbit-decay candidates.

\end{abstract}




\end{frontmatter}




\section{Introduction}
\label{sec:intro}









It is well known that transit observations have made significant contribution 
to the discoveries of extra-solar planets (exoplanets). 
With the transit method, 
ground-based telescopes were employed to do sky surveys and brought a chain of exoplanet 
detections \citep{Alonso+2004ApJ, Bakos+2004PASP, Pollacco+2006PASP}.  
Through the same method, space telescopes such as CoRoT (Convection, Rotation and planetary Transits), 
Kepler, Transiting Exoplanet Survey Satellite (TESS) also joined this
journey of discoveries \citep{Auvergne+2009AA, Borucki+2010Sci, Barclay+2018ApJS}. 

Because (a) the transit probability is higher for those planets closer to the host stars; 
(b) the transit event happens more frequently for planets with shorter orbital periods;
(c) the transit depth is larger if the planet size is bigger relatively to the host star,
a huge number of hot Jupiters have been detected. As these hot Jupiters are so close to their host stars,
theoretically, these planets might experience tidal orbital decay \citep{Jiang+2003ApJ}
or lose their masses through the mass transfer into Roche lobes \citep{Valsecchi+2015ApJ, Jackson+2016CeMDA}.

In order to further constrain the orbital parameters, and study the above theoretically predicted 
orbital decay or mass loss,  these discovered exoplanets' transit events at later epoches
have been monitored through many following-up observations intensively. For example, 
\citet{Winn+2009AJ} performed two transit
observations of the exoplanet WASP-4b. Their work led to the updated planetary mass and radius. They found that this planet was 15$\%$ larger than expected and was thus a bloated planet. 
\citet{Jiang+2013AJ} presented five transit 
light curves of TrES-3b and proposed that there could be a possible transit timing variation (TTV) for this planet. 
\citet{Mannaday+2020AJ} further studied this system and reconfirmed this possibility.
In addition, the possible orbital decay of WASP-43b \citep{Hellier+2011AA} was also investigated, and both positive and negative results 
were claimed controversially \citep{Blecic+2014ApJ,Jiang+2016AJ,Hoyer+2016AJ,Davoudi+2021AJ,Garai+2021MNRAS}.

Interestingly, \citet{Bouma+2019AJ} later employed 18 TESS transit light curves of WASP-4b and discovered that the transits
occurred earlier than previously predicted. They concluded that the orbital period of WASP-4b 
was changing, which could be caused by tidal orbital decay, apsidal precession, or the gravitational influence 
of a third body \citep{Miralda-Escude2002, A-thano+2022}.

In fact, TTV has also become one of the tools to detect new exoplanets. For example, 
a further analysis on transit timing data by \citet{Sun+2019} led to the discoveries of additional two new exoplanets in
Kepler-411 system, which was previously known to host two transiting planets.
Moreover, \citet{Barros+2022AA} employed 12 transit light curves to investigate the tidal deformation of WASP-103b
and \citet{Szabo+2022} found large-amplitude TTVs of AU Microcopii b and c through 
combined TESS and CHEOPS transit data. 

Finally, 
\citet{Maciejewski+2016AA} and \citet{Patra+2017AJ} collected mid-transit times of the exoplanet WASP-12b and confirmed the TTV of WASP-12b. They favored the model of orbital decay
and a decay rate of $-25.60\pm 4.0$ ms per year was reported. 
Further timing analysis \citep{Yee+2020ApJ, Turner+2021AJ, Wong+2022AJ} supports 
that the orbit of  WASP-12b is decaying.

On the other hand, the huge amount of light curves obtained by the Kepler Space Telescope 
leads to an excellent database of mid-transit times, which triggers many further studies. 
For example, \citet{Ford+2012ApJ} searched for TTVs from 
the first four-month Kepler data and studied multi-planet systems.
\citet{Steffen+2012ApJ} used statistics tests 
to search for TTVs among pairs of Kepler Object of Interest (KOI) in multi-transiting
systems. Finally, \citet{Holczer+2016ApJS} presented a transit timing catalog of 2599 KOIs
and showed clear TTVs among many multi-planet systems. Recently, these Kepler transit timing data were
employed by \citet{Wu+2023AJ} to search for hidden nearby companions to hot Jupiters through examining TTV signals,
and also used by \citet{Kipping+2023MNRAS} to search for exomoons.

In order to provide a platform to give information of transit events and collect transit light curves, 
Exoplanet Transit Database (ETD) was established \citep{Poddany+2010NewA}.
It allows both amateur and professional astronomers to upload 
transit light curves. It employs an analytic model to determine mid-transit times. 
The predicted information of coming transit events, the observational data of transit light curves, and values 
of mid-transit times are all publicly available from ETD website. \citet{Hagey+2022AJ} then took ETD mid-transit times
and searched for long-term period variations.
In addition, with a goal to constrain the future observational timing of proposed targets of ARIEL space telescope \citep{Pascale+2018}, 
\citet{Kokori+2021ExA} organized the ExoClock project which builds a big network of telescopes and obtain many exoplanet transit light curves. The ExoClock project also provides an 
interactive platform that light curves can be uploaded by observers and become publicly available. 
\citet{Kokori+2022ApJS} updated their telescope capabilities and reported their results of ephemerides of 180 planets. Recently, \citet{Kokori+20023ApJS}
further present a homogenous catalog of updated ephemerides for 450 planets.
The main objective of ExoClock is to continuously improve the precision of 
exoplanet ephemerides. 
After more mid-transit times are collected, the slope of the linear ephemerides is likely to be changed and the 
future transit timing would be drifted\citep{Kokori+2022ApJS}. However, if a linear function is still the best 
model for mid-transit times at different epoch, there is no TTV. The TTV could be confirmed to exist only when a 
non-linear model of mid-transit time gives a better fitting than linear models.

Moreover, \citet{Ivshina+2022ApJS} provided a database of transit times of 382 planets 
derived from TESS data.
With these huge amount of collected mid-transit-time data,
it is a good opportunity to search for possible orbital decays of hot Jupiters through data-model fitting.
In this paper, we collect all mid-transit times from \citet{Holczer+2016ApJS}, 
\citet{Ivshina+2022ApJS}, \citet{Kokori+20023ApJS} 
and focus on examining whether there are any suspected on-going orbital 
decays among transit hot-Jupiter planets.  

Therefore, we would firstly select hot Jupiters from all planets in these three catalogs.  
After data-model fitting, these hot Jupiters would be classified into several types based on the results of TTV analysis.
Among these, the most important planets are those belong to the orbit-decay-TTV type. 
With best-fit models, 
their values of stellar dissipation parameters or orbital evolution would be determined.   
This paper is organized as follows. \S 2 describes the mid-transit-time data and planet samples.
\S 3 describes the process of data-model fitting and presents the classification. 
\S 4 provides theoretical models and interpretations of the orbit-decay cases.  
\S 5 makes conclusions.

\section{The Mid-Transit-Time Data and Hot-Jupiter Samples}

\begin{figure*}
\centering
\includegraphics[width=0.7\linewidth]{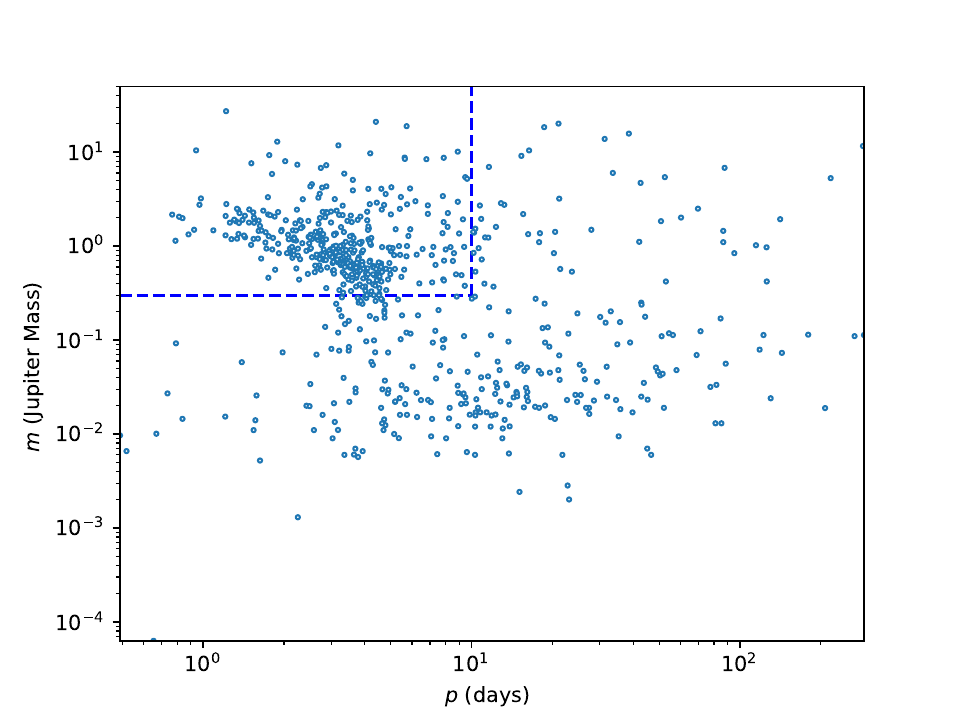}
\caption{The distribution of 700 planets on the $p-m$ plane.  
The dashed lines indicate the conditions of hot Jupiters.}
\label{fig:mpfig1}
\end{figure*}

In this paper, the mid-transit-time data from \citet{Holczer+2016ApJS} (Kepler Catalog), 
\citet{Ivshina+2022ApJS} (TESS Catalog), and \citet{Kokori+20023ApJS} (ExoClock Catalog)
are all collected. There are totally 2430 planets and the majority is from the Kepler Catalog. 
In order to estimate tidal dissipation parameters
and determine the orbital evolution, we focus on those exoplanets with known values of mass in this paper and leave those exoplanets with unknown values of mass to be studied in the future.
After examining the Extrasolar Planets Encyclopaedia (http://exoplanet.eu/), only 700 out of these 2430 planets
have the values of mass ($m$) and orbital period ($p$).
Fig. \ref{fig:mpfig1} shows the distribution of these 700 planets on the $p-m$ plane.  
After employing the definition of hot Jupiter \citep{ZhuWu2018AJ}, i.e. $p<10$ (days) and $m>0.3$ Jupiter-Mass,
390 planets are selected in the top-left corner of Fig. \ref{fig:mpfig1} as indicated by the dashed lines.  

In addition, for a particular planet, when there are more than one mid-transit-time data for the same transit epoch, 
we take the weighted mean to get the mid-transit time and the corresponding uncertainty for that epoch.  
In order to have sufficient mid-transit-time data points for further analysis, only those planets with 10 
or more mid-transit times are included in our study. 
Thus, only 309 out of 390 hot Jupiters will be taken to do the data-model fitting presented in the next section.


\section{The Classification}\label{secmodel}

Our goal is to examine whether there are any TTVs among the selected planets, i.e. hot Jupiters, in the data,
and whether these TTVs imply orbital decays. 
The minimization of chi-square functions would be performed and the corresponding reduced 
chi-square of best-fit models would be used as an indicator of good or bad fittings.
This is because the value of reduced chi-square is usually less than ten and
its meaning can be easily understood as an averaged weighted square of data-model differences.
In addition, the Bayesian Information Criterion (BIC) would be employed during the process
of selecting orbit-decay cases.

Based on these fitting results, exoplanets are classified into different categories.
We shall be aware that these calculations are based on statistics and we 
regard these results as likely tendencies with certain probabilities.

\subsection{The Data-Model Fitting}

For each planet, two analytic models would be employed to fit
the mid-transit time data through the emcee package \citep{Foreman+2013PASP},
which is an implementation of Goodman $\&$ Weare’s Affine Invariant Markov 
Chain Monte Carlo (MCMC) algorithm \citep{Goodman+2010CAMCS}. 
With the values of reduced chi-square of two different fitting models, we 
are able to classify these exoplanets into three categories, i.e.
the unclassified-TTV type, the null-TTV type, and the orbit-decay-TTV type.
The classification rule would be described in the later part of this subsection.
   
The first model corresponds to a fixed orbit with a 
constant orbital period, so the mid-transit time is a linear function of transit number.
That is
\beq
t_{tra}(N)=t_{f}+P_{f}N,
\label{eq:linear}
\eeq
where $t_{tra}$ is the mid-transit time as a function of $N$, $N$ is 
the transit number, $t_{f}$ is a reference time, and $P_{f}$ is the orbital period. 
There are two fitting parameters, $t_{f}$ and $P_{f}$, so 
the degree of freedom is the number of data point minus two for this model.  
After the data-model fitting process, we would obtain the best-fit $t_{f}$ and $P_{f}$
and the corresponding chi-square $\chi^2_f$. The value of reduced chi-square $R\chi^2_{f}$ is then
calculated as the chi-square divided by the degree of freedom. 
Through the MCMC sampling of data-model fitting,
the posterior probability distribution of fitting parameters can be obtained. 
Following \citet{HoggCraig1989} and using the standard deviation ( $\sigma$ ) 
of a normal distribution as the parameter uncertainty, 
the 15.9 percentile and 84.1 percentile of a posterior probability distribution are 
set as the uncertainty boundaries for all fitting parameters in this paper.

However, when the orbit of an exoplanet changes slowly, 
the first derivative of period with respect to $N$ 
will not be zero and the mid-transit time will be slightly
shorter or longer for each epoch. In this case, an exoplanet will drift inward or 
outward slightly every epoch. It is therefore worthwhile to consider a second model with more terms
to include possible orbital variations.

The second model is 
the one allows possible orbital variations \citep{Patra+2017AJ,Su+2021AJ} and
the mid-transit time $t_{tra}$ is a quadratic function of transit number $N$. That is
\beq
    t_{tra}(N)=t_{v}+P_{v}N+ \frac{1}{2}\frac{dP_{v}}{dN} N^2,
\label{eq:mig} 
\eeq
where  $dP_{v}/{dN}$  is the period derivative with respect to $N$. There are three 
parameters, $t_{v}$,$P_{v}$, and $dP_{v}/{dN}$.  The degree of 
freedom is the number of data point minus three in this model.
After the MCMC process, we will obtain the best-fit $t_{v}$, $P_{v}$, $dP_{v}/{dN}$, 
the chi-square's value $\chi^2_v$, and also the reduce 
chi-square's value which is denoted as $R\chi^2_{v}$. 

Please note that because we decide to focus on the orbital decays, 
we only consider those planets with the best-fit $dP_{v}/{dN}<0$.
After removing 165 planets with best-fit $dP_{v}/{dN}\geq 0$,
we finally have 144 hot Jupiters as our samples in this paper.

The values of best-fit reduced chi-square for above two models will be determined for all 144 hot Jupiters,
and thus all considered hot Jupiters have their own $R\chi^2_{f}$ and $R\chi^2_{v}$. 
When both best-fit reduced chi-square have values larger than or equal to 3.0, i.e. $R\chi^2_{f}\geq 3$ and $R\chi^2_{v}\geq 3$,
it means neither the fixed-orbit model nor orbit-decay model could fit the data well.
Thus, it indicates that, for this considered exoplanet, there are some level of TTV but the data might not closely follow
the prediction of our orbit-decay model. 
In this case, it is set to be the unclassified-TTV type.

For the rest of hot Jupiters, we need to determine whether they belong to the null-TTV type or the orbit-decay-TTV type.
Because the BIC has been the choice of methods used to pick up the better model in literature
\citep{Patra+2017AJ,Mannaday+2022AJ,Hagey+2022AJ},  we will calculate BIC values of 
both fixed-orbit model and orbit-decay model of a given hot Jupiter.
Once BIC values of two models are obtained, the standard way is to pick up the one with smaller BIC value as the preferred model
implied by the data.
However, the BIC values of two models could be very close for some cases, it is better to estimate the uncertainties
of BIC values through the bootstrap method \citep{Press+1992, Jiang+2007AJ}, in which
the BIC values are calculated for a large number of independent repetitions of bootstrap sampling.

Thus, for those planets with $R\chi^2_{f}< 3$ or $R\chi^2_{v} < 3$, we 
calculate the BIC values \citep{Patra+2017AJ,Mannaday+2022AJ,Hagey+2022AJ} of
the fixed-orbit model $BIC_f$ and the orbit-decay model $BIC_v$ as, 
\beq
\begin{array}{ll}
&BIC_f=\chi^2_f+k\log(n_m)  \\
&BIC_v=\chi^2_v+k\log(n_m), 
\end{array}
\eeq
where $\chi^2_f$ is the chi-square value of fixed-orbit model,  $\chi^2_v$ is the chi-square value of orbit-decay model,   
$n_m$ is the number of mid-transit-time data, and $k$ is the number of free
parameters. Here k=2 for the fixed-orbit model, and k=3 for the orbit-decay model. 

In order to estimate the uncertainties of $BIC_f$ and $BIC_v$, the bootstrap sampling is performed for 2000 times 
\citep{Jiang+2007AJ} and the 25, 50, 75 percentile values are determined.
The confidence intervals ($BIC_{f25},  BIC_{f75}$) and ($BIC_{v25},  BIC_{v75}$) are then obtained.
We set the planets with $BIC_{v75}<BIC_{f25}$ as the orbit-decay-TTV type,
and assign the rest to be the null-TTV type.

\subsection{The Unclassified-TTV Type}

After performing the above tasks, the classification can be done. 
There are 22 planets belonging to the unclassified-TTV type as listed in Table \ref{tab:unclass}.
Their values of 
$R\chi^2_{f}$, $R\chi^2_{v}$, and the number of mid-transit-time data are presented in Table \ref{tab:unclass}.
 
The orbital properties of these systems being listed in the unclassified-TTV type  
are not completely clear yet.
They deserve future further investigations.

\begin{table*}
\begin{center}
\begin{tabular}{|l|c|c|c||l|c|c|c|} \hline
Name & $R\chi^2_{f}$ & $R\chi^2_{v}$ & $n_m$  &Name & $R\chi^2_{f}$ & $R\chi^2_{v}$   & $n_m$ \\ \hline
CoRoT-11b &    11.10  &11.05 & 22 & WASP-16b  &   4.67 & 4.33 & 26 \\
CoRoT-2b &  61.45 & 58.45  & 138 &WASP-19b  &    4.64 & 4.27 & 145 \\
HAT-P-16b &   3.42 &3.44 &47 &WASP-2b &4.37 & 4.43 &61 \\
HAT-P-29b  &  9.66 & 10.04 & 20 &WASP-33b &    8.04 &8.35 & 29 \\
HAT-P-54b &    4.09 & 4.15 & 64 &WASP-41b &    3.76 & 3.87 & 39 \\
K2-237b & 4.69 & 4.71 & 35 &WASP-45b &
11.31 & 11.02 & 32 \\
KELT-10b & 4.02 & 3.17 & 11 &WASP-48b &    
4.06 & 4.09 & 111 \\
KELT-25b & 5.33 & 4.99 & 14 &WASP-56b  &
6.50 & 5.63 & 21 \\
KELT-4Ab & 5.16 &5.32 &25 &WASP-80b &     
4.47 & 4.07 & 32 \\
TrES-1b & 4.06 &3.42 & 81 & XO-3b&  6.60
&3.27 &48 \\
WASP-10b & 5.73 & 5.02 & 74 &XO-4b   & 3.81 
&3.62 &34 \\ \hline
\end{tabular}\end{center}
\caption{The values of $R\chi^2_{f}$, $R\chi^2_{v}$, and the number of mid-transit-time data 
of those planets belonging to the unclassified-TTV type.}
\label{tab:unclass}
\end{table*}

\subsection{The Orbit-Decay-TTV Type}

\begin{figure*}
\centering
\includegraphics[width=1\linewidth]{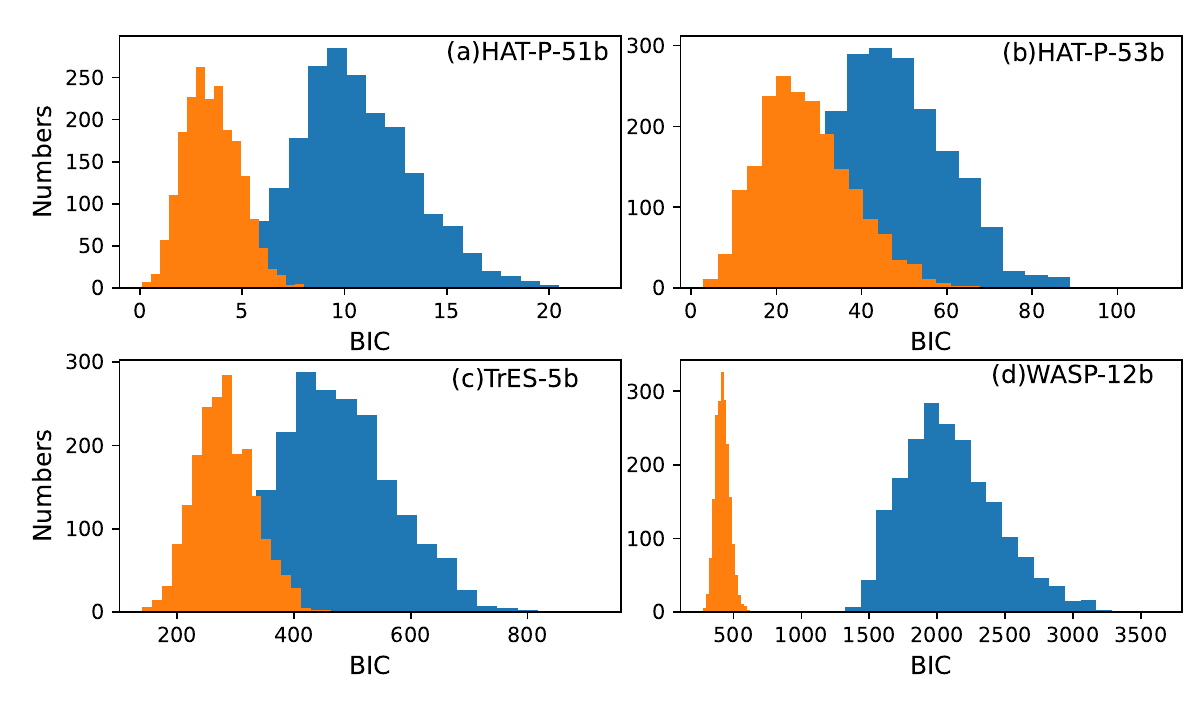}
\caption{The histograms of BIC values for the four planets being the orbit-decay-TTV type.  
The BIC values of the fixed-orbit model are in blue color, and the 
BIC values of the orbit-decay model are in orange color.}
\label{fig:hisBIC}
\end{figure*}

After the bootstrap sampling, the histograms of BIC values of all those planets not in unclassified-TTV type
are obtained and their corresponding confidence intervals ($BIC_{f25},  BIC_{f75}$) and ($BIC_{v25},  BIC_{v75}$) are then 
determined. 
Fig. \ref{fig:hisBIC} presented the BIC histograms of four planets with  $BIC_{v75}<BIC_{f25}$. Their 
BIC values of orbit-decay model are smaller than the values of fixed-orbit model at the 75\% confidence level. 
Thus, they are set as the orbit-decay-TTV type.

In Table \ref{tab:bf_orb}, 
the names of planets, the values of $R\chi^2_{f}$, $R\chi^2_{v}$, the best-fit parameters
$P_v$, $t_v$, $dP_{v}/dN$, and the number of mid-transit-time data are listed.  
 
\begin{table*} 
\centering
\begin{tabular}{|l|l|l|l|l|l|l|} \hline
Name &$R\chi^2_{f}$&  $R\chi^2_{v}$ & $P_{v}$ (days) & $t_{v}$ (BJD) & $dP_{v}/{dN}$ &$n_m$ \\ 
\hline
HAT-P-51b &   1.31 & 0.50 & $4.218028\pm 3\times 10^{-6}$ & $2456194.1228 \pm 0.0004$ & $-1.80\times 10^{-8} \pm 7\times 10^{-9}$ & 10 \\
HAT-P-53b & 3.10 & 1.92 & $1.961628 \pm 1\times 10^{-6}$ & $2455829.4483 \pm 0.0004$ & $-5.1\times 10^{-9} \pm 1\times 10^{-9}$  & 17  \\    
TrES-5b &  4.60 & 2.75 &  $1.4822490 \pm 2\times 10^{-7}$ & $2455152.7316 \pm 0.0001$ & $-1.5 \times 10^{-9} \pm 1\times 10^{-10}$  & 105  \\  
WASP-12b  & 7.46 & 1.50 & $1.09142185  \pm 7\times 10^{-8}$  & $2454508.97693\pm 7\times 10^{-5}$ & $ -1.05 \times 10^{-9}\pm 3\times 10^{-11}$  & 283         \\ 
\hline
\end{tabular}
\caption{The values of $R\chi^2_{f}$, $R\chi^2_{v}$, 
the best-fit parameters of the orbit-decay model, and the number of mid-transit-time data 
of four planets belonging to the orbit-decay-TTV type.}
\label{tab:bf_orb}
\end{table*} 

\begin{figure*}
\centering
\includegraphics[width=1\linewidth]{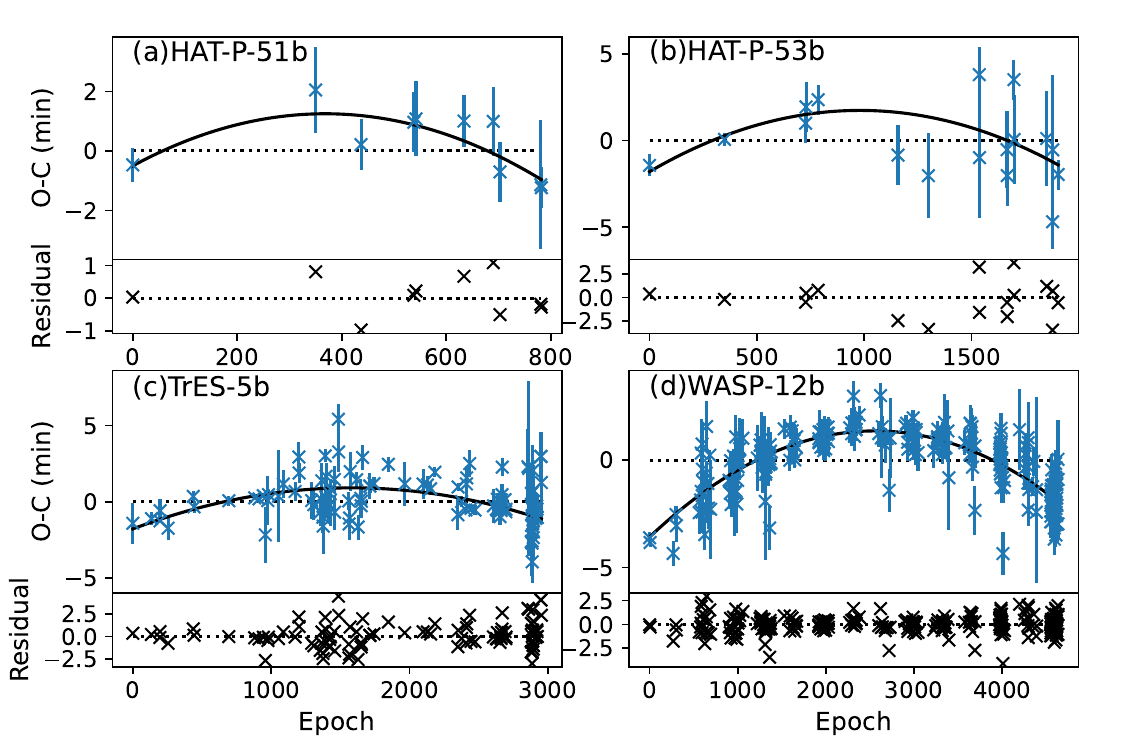}
\caption{The O-C diagrams of four planets belonging to the orbit-decay-TTV type }
\label{fig:OCdec}
\end{figure*} 

In Fig. \ref{fig:OCdec}, the O-C diagrams of those planets being the orbit-decay-TTV type are presented.
Both HAT-P-51b and HAT-P-53b were discovered by \citet{Hartman+2015AJ}.  HAT-P-51b is a 0.309
Jupiter-Mass planet with orbital period 4.2 days
while HAT-P-53b is a 1.487 Jupiter-Mass planet with 
orbital period 1.96 days.
TrES-5b was discovered by \citet{Mandushev+2011ApJ}. It is a hot Jupiter
with mass 1.778 Jupiter-Mass and an orbital period 1.48 days. 
WASP-12b is a 1.47 Jupiter-Mass planet with orbital period 1.09 days. It is a best-known 
orbit-decay exoplanet. According to our best-fit values of $dP_{v}/dN$, HAT-P-53b, TrES-5b, and WASP-12b have similar order of 
orbit-decay rates but HAT-P-51b has a 
larger orbit-decay rate.
Their star-planet tidal interactions and the corresponding 
dissipation parameters will be further discussed in the next section.

\subsection{The Null-TTV Type}

For the null-TTV type, 
there are 118 planets. 
The values of of $R\chi^2_{f}$, $R\chi^2_{v}$, the 
best-fit values of $P_{f}$, $t_{f}$ in Eq. (\ref{eq:linear}), which lead to new ephemerides, 
and the number of mid-transit-time data
are presented in Table \ref{tab:linear}.  

\begin{table*}
\centering
\begin{tabular}{|l|l|l|l|l|l|} \hline
Name &$R\chi^2_{f}$&  $R\chi^2_{v}$  & $P_{f}$ (days) & $t_{f}$ (BJD)  & $n_m$ \\ \hline
CoRoT-19b  &   0.97 & 0.99 &
$3.897138\pm 2\times 10^{-6}$ & $2455261.3388 \pm 0.0005$ & 19    \\      
CoRoT-5b  &   0.27 & 0.27 & $4.037916\pm 1\times 
10^{-6}$ & $2454400.1990\pm 0.0002$ &  34   \\   
EPIC246851721b & 1.89 & 2.01 & $6.180268 \pm 2\times 10^{-6}$ & $2457828.0285 \pm 0.0002$ &17 \\         
HAT-P-1b & 1.34 & 1.27 & $4.4652991\pm 4\times 10^{-7}$ & $2453979.9322 \pm 0.0002$ & 25 \\     
HAT-P-13b  &   2.20 & 2.25 & $2.9162426 \pm 4\times 10^{-7}$ & $2454581.6256 \pm 0.0002$ & 47 \\         
HAT-P-2b  &  1.32 & 1.41 & $5.6334686\pm 5\times 10^{-7}$  & $2454212.8566 \pm 0.0004$ &  14  \\  
HAT-P-22b & 1.71 & 1.63 & $3.2122321 \pm 2\times 10^{-7}$ & $2454930.2212\pm 0.0002 $ &  47 \\    
HAT-P-23b &   1.51 & 1.53 & $1.21288643 \pm 5\times 10^{-8}$ & $2454632.7329\pm 0.0001$ & 79 \\
HAT-P-24b &   1.70 & 1.69 & $ 3.3552444 \pm 
 2\times 10^{-7}$ & $2455206.9115 \pm 0.0002$ & 42 \\        
HAT-P-25b  &  1.63 & 1.58 & $3.6528152 \pm 3\times 10^{-7}$ & $2455136.6717 \pm 0.0002$ &46 \\         
HAT-P-27b  &  1.44 & 1.31 & $3.0395780 \pm 2\times 10^{-7}$ &$2455186.0203 \pm 0.0002$  &24 \\ 
......&......&.....&  ....... &  ........&....\\
XO-7b   &    1.45 & 1.47  & $2.86413530\pm 8\times 10^{-7}$ & $2457917.4754\pm 0.0003$ & 47 \\  
\hline
\end{tabular}
\caption{The values of $R\chi^2_{f}$, $R\chi^2_{v}$, 
the best-fit parameters of the fixed-orbit model, and the number of mid-transit-time data 
of those planets belonging to the null-TTV type. This table is available in its entirety in machine-readable form.
A portion is shown here for guidance regarding its form and content.
}
\label{tab:linear}
\end{table*}

\section{The Systems with Orbital Decays}

In order to investigate
the physics of orbital evolution of the star-planet systems with orbital decays, theoretical models of 
orbital decays are described here.
Following \citet{Goldreich+1966Icar}, \citet{Jackson+2008ApJ}, and \citet{Murray+1999}, 
the orbital evolution of a planet influenced by the tidal effect can be described as 
\beqn
\frac{1}{a}\frac{da}{dt}&=&-\left[\frac{63}{2}\left(GM_\ast^3\right)^{1/2}\frac{R_p^5}{Q_p M_p}e^2\right] a^{-13/2}\non\\
&&+{\rm sign}\,\,(\Omega-n)\frac{9}{2}\left(\frac{G}{M_\ast}\right)^{1/2}\frac{R_\ast^5M_p}{Q_\ast} a^{-13/2},\label{eq:dadt1}\\
\frac{1}{e}\frac{de}{dt}&=&-\left[\frac{63}{4}\left(GM_\ast^3\right)^{1/2}\frac{R_p^5}{Q_p M_p}\right] a^{-13/2}\non\\
&&+{\rm sign}\,\,(2\Omega-3n)\frac{171}{16}\left(\frac{G}{M_\ast}\right)^{1/2}\frac{R_\ast^5M_p}{Q_\ast} a^{-13/2}, 
\label{eq:dedt1}
\eeqn
\no where $a$ is the orbital semi-major axis, $e$ is the orbital eccentricity, $t$ is the time, 
$G$ is the gravitational constant, $R$ is a body's radius, $M$ its mass and $Q$ its tidal dissipation parameter, and subscripts p and
$\ast$ refer to the planet and star, respectively. 
In addition, $\Omega$ is the axial angular velocity of star and $n$ is the mean motion of the planet.
These two equations determine the time derivatives of orbital semi-major axis and eccentricity.
For the case that orbital eccentricity is not equal to zero, the above two equations are coupled 
and need to be solved simultaneously. 
For the case of circular orbits, i.e. eccentricity $e=0$, only the 2nd term of Eq.(\ref{eq:dadt1})
needs to be considered. 
To solve the above equations numerically, some
planetary and stellar parameters are needed and thus listed in Table \ref{tab:para_starplanet} for convenience.

\subsection{The Planets with Circular Orbits}

Among four orbit-decay cases, TrES-5b and WASP-12b
are those two planets moving on circular orbits.
Under the assumption that $e=0$ and $\Omega<n$ \citep{Penev+2018, Matsumura+2010}, 
from Eq.(\ref{eq:dadt1}), we have 
\beq
\frac{da}{dt}=-\frac{9}{2}\left(\frac{G}{M_\ast}\right)^{1/2}\frac{R_\ast^5M_p}{Q_\ast} a^{-11/2}.\label{eq:dadt2}
\eeq
Since the orbital period $P_v$ satisfies $P^2_v=4\pi^2a^3/\mu$ and $\mu=G(M_\ast+M_p)\sim GM_\ast$, we substitute this relation into 
Eq.(\ref{eq:dadt2}) and have 
\beq
\frac{1}{P_v}\frac{dP_v}{dN}=\frac{dP_v}{dt}=\frac{3\pi}{\sqrt{GM_\ast}}a^{1/2}\frac{da}{dt}
=-\frac{27\pi}{2Q_\ast}\left(\frac{M_p}{M_\ast}\right)\left(\frac{R_\ast}{a}\right)^5.
\label{eq:q_ast}
\eeq
The above equation, i.e. Eq.(\ref{eq:q_ast}),  gives the relation between
$Q_\ast$ and $dP_v/dN$. This relation enables us to estimate the stellar dissipation parameter
$Q_\ast$ from transit observations.
In principle, $Q_\ast$ is inversely proportional to $|dP_v/dN|$.

The data-model fitting in the previous section leads to the best-fit $dP_v/dN$ and
also the posterior distribution of $dP_v/dN$. 
The best-fit values of $Q_\ast$ are obtained and shown at the 2nd column of Table \ref{tab:para_od}.
In addition, using the 2.3 and 97.7 percentile values 
(i.e. $2\sigma$) of the posterior distribution of $dP_v/dN$, 
the lower limit and upper limit of $Q_\ast$ are also obtained and shown 
at the 3rd and 4th column of Table \ref{tab:para_od}. They are denoted as $LQ_\ast$ and $UQ_\ast$. 


\begin{table*} 
\centering
\begin{tabular}{|l|l|l|l|l|l|} \hline
Name & $M_p(M_J)$ & $a(AU)$  &  $M_\ast(M_\odot)$ & $R_\ast(R_\odot)$ & e \\ 
\hline
 HAT-P-51b &  0.309 &  0.05069  &   0.976 &     1.041 &0.123 \\
HAT-P-53b & 1.487   & 0.03159   & 1.093 &    1.209 & 0.134 \\ 
 TrES-5b  & 1.778 &    0.02446  & 0.893 &    0.866  & $0^{[a]}$ \\
WASP-12b  &1.47      &0.02344  & 1.434   &  1.657& 0  \\  \hline
\end{tabular}
\caption{The planetary mass, planetary orbital semi-major axis, star's mass, star's radius, and planetary orbital eccentricity 
for planets belonging to the orbit-decay-TTV type. These parameters are adopted from {\it The Extrasolar Planets Encyclopaedia http://exoplanet.eu/}. 
Note [a] : This eccentricity value is from \citet{Mandushev+2011ApJ}. 
}
\label{tab:para_starplanet}
\end{table*} 

\begin{table*} 
\centering
\begin{tabular}{|l|r|r|r|} \hline
 Name  & $Q_\ast$ & $LQ_\ast$ & $UQ_\ast$\\ 
\hline
HAT-P-51b  &   - &  -& - \\
HAT-P-53b  &   - & - & - \\
TrES-5b &     $9619.98^{+740.04}_{-645.92}$ &  8412.426 &  12324.42 \\
WASP-12b  &   $165140.76^{+4134.33}_{-3930.96}$ &  157452.63 & 178548.32 \\ \hline
\end{tabular}
\caption{The best-fit stellar dissipation parameter $Q_\ast$, its lower limit $LQ_\ast$, and upper limit $UQ_\ast$
for planets belonging to 
the orbit-decay-TTV type. Note that there are no determined values for HAT-P-51b and HAT-P-53b as
their eccentricities are not zero.
}
\label{tab:para_od}
\end{table*}

The result here that TrES-5b being the orbit-decay-TTV type is actually consistent with several previous studies. For example, \citet{Sokov+2018}
proposed the possible existence of an additional planet in order to explain the 
detected TTV. \citet{Maciejewski+2021A&A} revisited this system by providing 
more transit light curves. They did not confirm 
the existence of additional planet, but concluded that the orbital period of TrES-5b
could vary on a long timescale.
Employing ETD data, \citet{Hagey+2022AJ} 
gave a list of systems, including TrES-5b, which have deviations from a constant orbital period.
Our best-fit $Q_\ast$ is around $10^4$ which is close
to the lower end of the suggested range in 
\citet{Matsumura+2010}.

WASP-12b is a well-known confirmed orbit-decay case \citep{Patra+2017AJ}.  
From Table \ref{tab:para_od}, the value of $Q_\ast$ is about $1.65\times 10^5$.  
Previously, \citet{Patra+2017AJ} estimated $Q_\ast$ to be about $2\times 10^5$ and 
\citet{Maciejewski+2018AcA} reported a value of $Q_\ast=(1.82\pm 0.32)\times 10^5$.
Thus, our result is consistent with previous results and we here reconfirm this orbit-decay case.

\subsection{The Planets with Eccentric Orbits}

As shown in \citet{Penev+2018, Matsumura+2010}, the stellar spins are usually slower than the orbital motion of hot Jupiters,   
we thus assume that the axial angular velocity of star is smaller than the mean motion of the planet, i.e. $\Omega < n$.
From Eq.(\ref{eq:dadt1}) and Eq.(\ref{eq:dedt1}), we have
\beqn
\frac{1}{a}\frac{da}{dt}&=&-\frac{63}{2}\left(GM_\ast^3\right)^{1/2}\frac{R_p^5}{Q_p M_p}e^2 a^{-13/2}\non\\
&&+\frac{9}{2}\left(\frac{G}{M_\ast}\right)^{1/2}\frac{R_\ast^5M_p}{Q_\ast} a^{-13/2}, \label{eq:dadt3}
\eeqn

\beqn
\frac{1}{e}\frac{de}{dt}&=&-\frac{63}{4}\left(GM_\ast^3\right)^{1/2}\frac{R_p^5}{Q_p M_p} a^{-13/2}\non\\
&&+\frac{171}{16}\left(\frac{G}{M_\ast}\right)^{1/2}\frac{R_\ast^5M_p}{Q_\ast} a^{-13/2}.\label{eq:dedt3}
\eeqn

In order to determine the evolution of semi-major axis $a$ and orbital eccentricity $e$ numerically from Eq.(\ref{eq:dadt3}) and Eq.(\ref{eq:dedt3}),
we need to set the values of stellar dissipation parameter $Q_\ast$ and planetary dissipation parameter $Q_p$.
As mentioned in \citet{Essick+2016ApJ}, the value of parameter $Q_\ast$ also depends on the shape of orbit and the mass of companion.
It is related to energy dissipation through heat and waves generated by tidal forces. 
For solar-type binary stars, it is about $10^6$ \citep{Meibom2005ApJ}.
For hot Jupiters, \citet{Essick+2016ApJ} suggested the value of $Q_\ast$ to be about  $10^5\sim 10^6$ 
but \citet{Matsumura+2010} considered a larger range as $10^4\sim 10^9$.
As for the planetary dissipation parameter $Q_p$, 
\citet{Trilling+2000ApJ} suggested a value $Q_p = 10^5$, but \citet{Jackson+2008ApJ} gave $Q_p=10^{6.5}$. 

\begin{figure*} 
\centering
\includegraphics[width=1\linewidth]{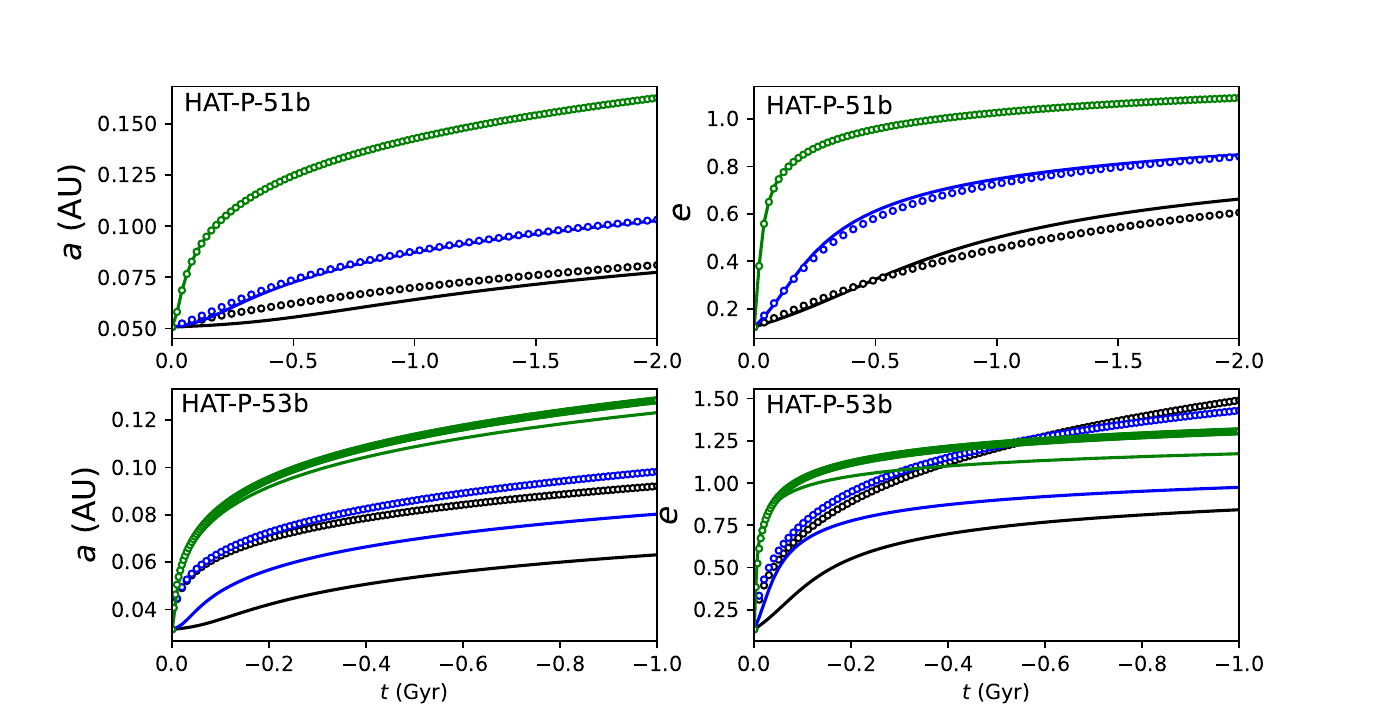}
\caption{The semi-major axis $a$ and eccentricity $e$ as functions of the backward time for HAT-P-51b and HAT-P-53b. In each panel, 
circles are for $Q_\ast=10^4$;
curves are for $Q_\ast=10^9$;
the green color is for $Q_p=10^5$;
the blue color is for $Q_p=10^6$;
the black color is for $Q_p=10^{6.5}$.
}
\label{fig:atet}
\end{figure*}

To show the effect of different values of the parameters $Q_\ast$ and $Q_p$,  we set two values of $Q_\ast$, i.e. $10^4$ and $10^9$,
and three values of $Q_p$, i.e. $10^5$, $10^6$, $10^{6.5}$.
With these values, Eq.(\ref{eq:dadt3})-(\ref{eq:dedt3}) are integrated backward in time \citep{Jackson+2008ApJ} and the
results are presented in Fig.\ref{fig:atet}.
There are six cases in each panel, in which
different marks are for different $Q_\ast$ values
and different colors are for different $Q_p$ values.
That is, circles are for $Q_\ast=10^4$;
curves are for $Q_\ast=10^9$;
the green color is for $Q_p=10^5$;
the blue color is for $Q_p=10^6$;
the black color is for $Q_p=10^{6.5}$.

The horizontal axes show the backward time with $t=0$ for the current time.  
Considering the evolution backward in time, 
the semi-major axis $a$ and eccentricity $e$
increase quickly 
from $t=0$ to $t=-0.1$ Gyrs 
and then increase slowly for both HAT-P-51b 
and HAT-P-53b.
We can also find that the result of HAT-P-51b 
does not really depend on the values of $Q_\ast$.
The result is mainly influenced by the values of $Q_p$. However, the result of HAT-P-53b are 
influenced by both $Q_\ast$ and $Q_p$.

\begin{figure*} 
\centering
\includegraphics[width=0.8\linewidth]{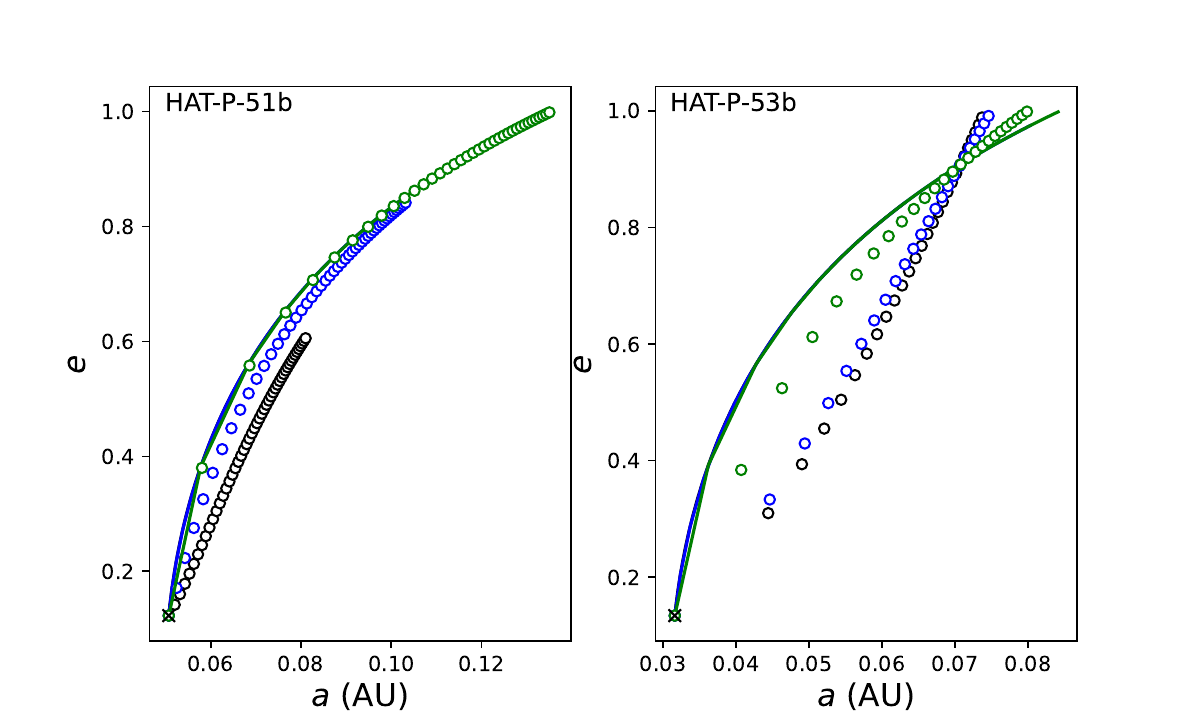}
\caption{The orbital evolution on the $a-e$ plane for HAT-P-51b and HAT-P-53b. In each panel, 
circles are for $Q_\ast=10^4$;
curves are for $Q_\ast=10^9$;
the green color is for $Q_p=10^5$;
the blue color is for $Q_p=10^6$;
the black color is for $Q_p=10^{6.5}$.
The crosses indicate the current values of 
semi-major axis $a$ and eccentricity $e$. 
}
\label{fig:ae}
\end{figure*}

It is obvious that the eccentricity goes
beyond $e=1$ for many cases, which are unlikely to be physical.  
For HAT-P-51b, the cases 
with $Q_p=10^6$ and $Q_p=10^{6.5}$ are physical.
For HAT-P-53b, only when $Q_\ast=10^9$, the
cases with $Q_p=10^6$ and $Q_p=10^{6.5}$ are physical.
Ignoring those parts of evolution
with eccentricity beyond $e=1$, 
we can collect the results within $0<e<1$ 
of all six cases 
for both HAT-P-51b and HAT-P-53b and see the 
evolution on the $a-e$ plane, as presented
in Fig.\ref{fig:ae}.
These theoretical models show that the 
semi-major axis $a$ and eccentricity $e$ 
would continue to decrease monotonically 
in the future.

\section{Conclusions}

The transit observations continuously make contributions to the exoplanet research. 
It does not only bring the discoveries of new systems, but also characterize their orbital evolution.  
The presented observational results here are directly derived from 
the combination of Kepler Catalog \citep{Holczer+2016ApJS}, 
TESS Catalog \citep{Ivshina+2022ApJS}, 
and ExoClock Catalog \citep{Kokori+20023ApJS}. This combination forms the most complete database up to date. 
In order to focus on those exoplanets which
could be influenced by the star-planet tidal interactions, the planets with 
orbital period $p<10$ days and planetary mass
$m>0.3$ Jupiter-Mass are picked as hot Jupiters. 
After performing the data-model fitting with both fixed-orbit and orbit-decay models,
only those with best-fit $dP_v/dN<0$ become the sample planets here.

Among these 144 planets, 118 are classified as the null-TTV type, 22
belong to the unclassified-TTV type, and 4 are classified as the orbit-decay-TTV type. 
Explicitly, 
HAT-P-51b, HAT-P-53b, TrES-5b, WASP-12b
are the orbit-decay cases.
Therefore,
our results reconfirm that WASP-12b is an orbit-decay planet, and also show that 
HAT-P-51b, HAT-P-53b, TrES-5b are potential
orbit-decay candidates. Apparently, 
future observational data would be very helpful to 
reveal the orbital nature of HAT-P-51b, HAT-P-53b, TrES-5b, 
as well as 
those 22 unclassified-TTV-type planets.

\section*{Acknowledgements}
We are grateful to the anonymous referee for good suggestions, which help
to improve this paper significantly.
This project is supported in part 
by the National Science and Technology Council, Taiwan, under
Li-Chin Yeh's 
Grant MOST 111-2115-M-007-008
and Ing-Guey Jiang's
Grant MOST 111-2112-M-007-035.




\bibliographystyle{elsarticle-harv} 
\bibliography{main}






\end{document}